\documentclass[conference]{IEEEtran}
\IEEEoverridecommandlockouts
\usepackage{cite}
\usepackage{amsmath,amssymb,amsfonts,flushend}
\usepackage{hyperref}
\usepackage{algorithmic}
\usepackage{graphicx}
\usepackage{textcomp}
\usepackage{xcolor}
\def\BibTeX{{\rm B\kern-.05em{\sc i\kern-.025em b}\kern-.08em
    T\kern-.1667em\lower.7ex\hbox{E}\kern-.125emX}}
\begin{document}

\title{DRL-based Slice Placement under Realistic Network Load Conditions}

\author{\IEEEauthorblockN{José Jurandir Alves Esteves\IEEEauthorrefmark{1}\IEEEauthorrefmark{2}, Amina Boubendir\IEEEauthorrefmark{1}, Fabice Guillemin\IEEEauthorrefmark{1} and Pierre Sens\IEEEauthorrefmark{2}}
\IEEEauthorblockA{\IEEEauthorrefmark{1}Orange Labs, France} \IEEEauthorrefmark{2}Sorbonne Universit\'e / CNRS / Inria, LIP6, France\\ \{josejurandir.alvesesteves, amina.boubendir, fabrice.guillemin\}@orange.com, pierre.sens@lip6.fr }

\maketitle

\begin{abstract}

We propose to demonstrate a network slice placement optimization solution based on Deep Reinforcement Learning (DRL), referred to as  Heuristically-controlled DRL, which uses a heuristic to control the DRL algorithm convergence. The solution is adapted to realistic networks with large scale  and under non-stationary traffic conditions (namely, the network load). We demonstrate the applicability of the proposed solution and its higher and stable performance over a non-controlled DRL-based solution. Demonstration scenarios include  full online learning with multiple volatile network slice placement request arrivals.   

\end{abstract}


\section{Context and Motivation}

%%Define acronyms in this section: ILP, PSN, NSPR. 

The big promise of Network Slicing is  to enable a high level of service customization in 5G networks and beyond. With the adoption of Network Slicing, telecommunications networks will become programmable platforms capable of offering virtual networks enriched by Virtual Network Functions (VNFs) and IT resources tailored to the specific needs of certain customers \cite{3GPP,etsi}. This paper is motivated by the  management issues raised by Network Slicing and more specifically by network slice placement. 

It is well known in the technical literature that network slice placement is  an $\mathcal{NP}$-hard optimization problem \cite{vne_np_hardness} that consists of choosing  which servers of the Physical Substrate Network (PSN) should host the VNFs composing a Network Slice  and which paths to use to steer traffic between these VNFs \cite{survey_vnf_ra_2016}. Some papers about network slice placement have recently used Deep Reinforcement Learning (DRL) in order to solve this problem in a scalable way \cite{p1, HA_DRL_TNSM}. However, most of these existing works applying DRL  assume a stationary environment, i.e., with a static network load. 

In real networks, traffic conditions are basically non-stationary with daily and weekly variations and subject to drastic changes like traffic peaks due to unpredictable events. This generates difficulties for the DRL algorithm to properly learn the environment to place slices. As a matter of fact, the continuously  changing network environment and policies may not be aligned with the algorithm previously acquired knowledge to find optimal solutions. %By consequent, the usage of the DRL algorithm in a online fashion can then become impractical.

This paper describes the Proof-of-Concept (PoC) of a detailed paper to be presented at \cite{cnsm_2021}. It presents a demonstration of the controlled DRL-based  network slice placement. We give in the present document an insight into our Heuristically Assisted DRL (HA-DRL) approach to network slice placement tailored to cope with the traffic changes. The proposed demonstration specifically addresses a large scale network scenario in which we apply HA-DRL in a fully online learning mode with time-varying network loads to show its advantages over a state-of-the-art pure-DRL approach.

The paper is organized as follows. First, we highlight the  objectives in Section~\ref{demo_overview}. Then, we present the solution architecture in Section~\ref{demo_architecture} and detail its implementation in Section~\ref{demo_implementation}. The planned demonstration is described in Section~\ref{demo_planned_demo}. We conclude the paper in Section~\ref{conclusion}.

\section{Slice Placement Under Realistic Network Load Conditions: PoC Architecture}

\subsection{Objective and Overview \label{demo_overview}}

The objective of this PoC is to show how the proposed network slice placement algorithm presented in the associate full paper \cite{cnsm_2021} is running. We precisely analyse the capacity of this solution to providing accurate placement decisions with small execution time and to achieve a high slice acceptance ratio in an online optimization environment, where Network Slice Placement Requests (NSPRs) are not known in advance and their demands are volatile; slices requests  arrive with a random rate and slices  stay in the PSN for a random duration. 

We evaluate four different placement algorithms fully described in \cite{cnsm_2021}: 
\begin{itemize}
    \item \textbf{DRL:} It is the pure-DRL algorithm we initially proposed in \cite{HA_DRL_TNSM}. The state representation does not include network load features and it does not use heuristic convergence control;
    \item \textbf{eDRL:} This algorithm is an enhanced version of DRL in which the state representation includes network load features. But, this algorithm  does not use heuristic convergence control; 
    \item \textbf{HA-DRL:} This algorithm uses heuristic convergence control but the state representation does not include network load features; 
    \item \textbf{HA-eDRL:} This algorithm state representation includes the network load  features and uses heuristic convergence control. 
\end{itemize}

\begin{figure*}[t] 
\centering
\includegraphics[width=0.9\linewidth] {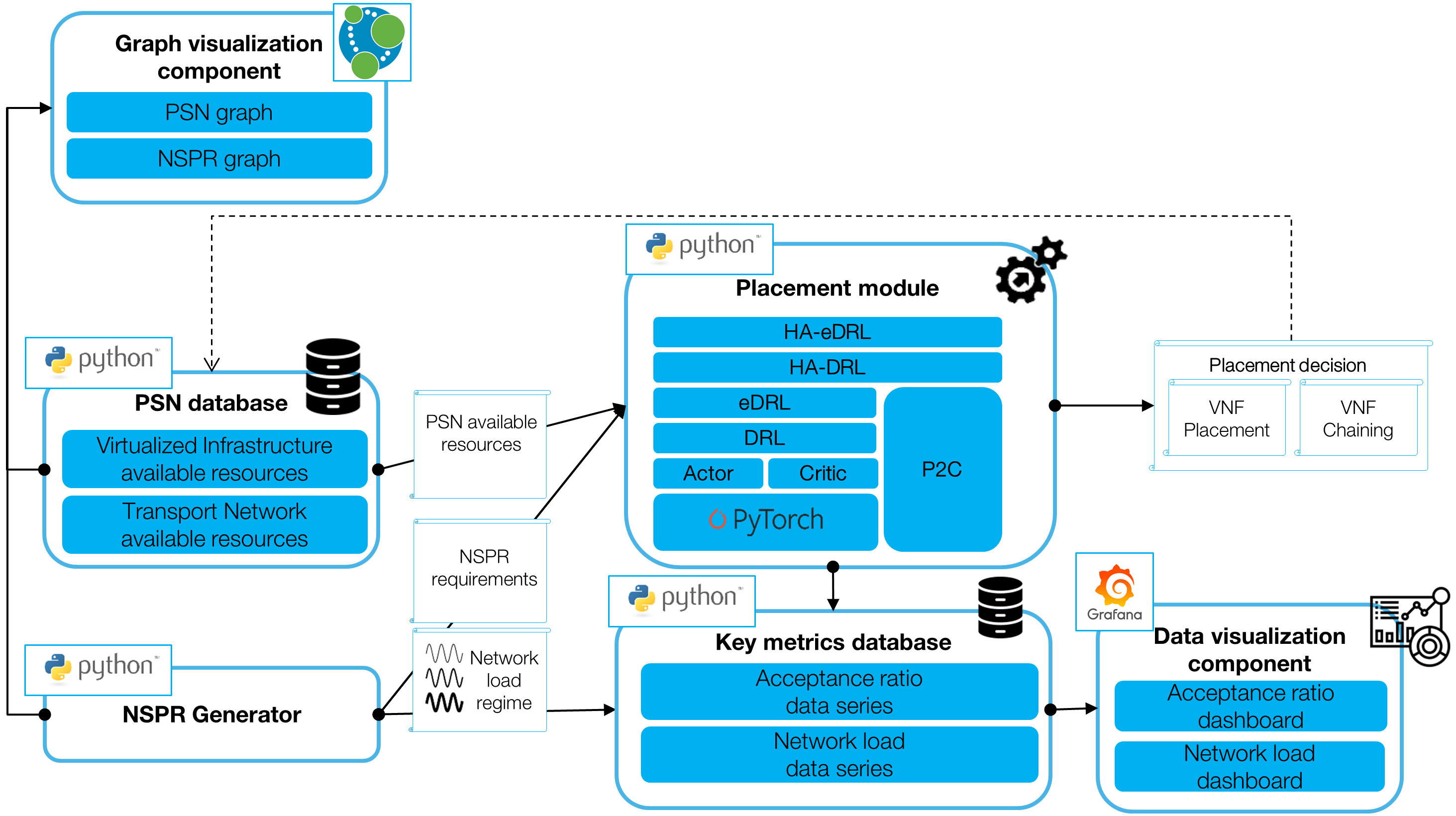}
\caption{Proof-of-Concept  Architecture: Optimal Network Slice Placement Under Non-stationary Traffic Conditions}
\label{fig:archi}
\end{figure*}

All algorithms are designed to optimize network slice acceptance ratio, resource consumption and load balancing under constrained PSN capacities (CPU/RAM hosting nodes capacities and bandwidth capacities on the links) and NSPR requirements (CPU/RAM VNFs requirements, bandwidth and requirements on the virtual links). 

\subsection{Proof-of-Concept Architecture Description \label{demo_architecture}}

The architecture of our Network Slice Placement scheme is given in the Fig.~\ref{fig:archi}. The NSPR generator is in charge of generating NSPR arrivals according to a dynamic network load regime. The PSN database stores the data about the available resources of the PSN. NSPR requirements and PSN available resources data are used as input by the Placement module. This latter implements the DRL-based algorithms and also the Power of Two Choices (P2C) heuristic algorithm \cite{cnsm_2020} used by HA-DRL and HA-eDRL algorithms to accelerate convergence. 

All algorithms calculate: i) a VNF placement decision, that is, where each VNF of the NSPR is to be placed and ii) a VNF chaining decision, that is, which paths in the network to use to interconnect the different VNFs. The Placement module can be configured to use one of the Placement algorithms  or both if comparison of Placement solutions is necessary. We compare the performance of the different algorithms in our demonstration.

Once the calculation of the Placement decision is done for one NSPR, an update of the available resources on the PSN is made and some key performance metrics are registered in the form of data series; the key metrics are: the acceptance ratio of network slices and the resource usage. These time series are used by the Data visualization component to build two dashboards: an acceptance ratio dashboard and a network load dashboard. Both dashboards are used to show the performance of the algorithms according to variations on the network load in real time. Finally, the  graph visualization component is used to allow the visualisations of the PSN and NSPR graphs.

\subsection{Solution Implementation Description \& Used Tools \label{demo_implementation}}

We have implemented the proposed network slice placement solution and we describe below the different tools used to implement its different components:
\begin{itemize}
    \item \textbf{Python:} We use  Python  for implementing different elements of the proposed solution:
    \begin{enumerate}
        \item the PSN database, that is actually represented by a series of data frames loaded from csv files,
        \item the NSPR generator that is actually a function that receives some parameters representing the CPU, RAM and bandwidth requirements of the NSPRs to be generated and also the parameters for the non-stationary network load model described in the detailed paper \cite{cnsm_2021},
        \item the placement algorithms.
         \end{enumerate}
          The DRL and eDRL algorithms are built upon an implementation of the  Actor and Critic Deep Neural Networks (DNNs) using the PyTorch framework. The HA-DRL and HA-eDRL algorithms are implemented on top of these DNNs and also of different classes and functions used to implement the P2C heuristic designed and implemented from scratch using only default Python packages.
    \item \textbf{Neo4j:} A Neo4j graph database  represents and  displays the PSN graph and the NSPR graph together with its requirements.
    \item \textbf{MySQL:} We use the MySQL database manager system to implement the Key metrics database with one table for the Acceptance ratio data series and another one for the Network load data series.
    \item \textbf{Grafana:} We use the Grafana tool to implement the Data visualization component in which we represent two dashboards using  the MySQL database of Key metrics as data-source. 
\end{itemize}

\section{Planned Demonstration \label{demo_planned_demo}}

We perform an emulation of the network slice placement decision making process using our implementation. 

\subsection{Considered Demonstration Scenarios}

We emulate a realistic PSN topology with 21 data centers (DCs) with different resource capacities. 

\begin{figure}[htbp] 
\centering
\includegraphics[width=0.9\linewidth] {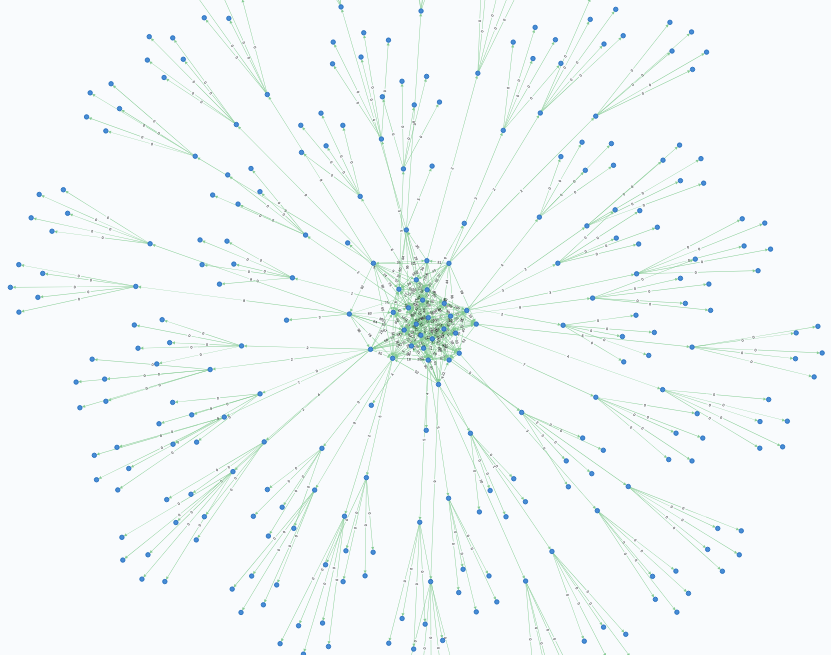}
\caption{Physical Substrate Network view using the Graph-based Visualization Component}
\label{fig:neo4j-psn}
\end{figure}

We consider 15 Edge Data Centers (EDCs) as local DCs with small resources capacities, 5 Core Data Centers (CDCs) as regional DCs with medium resource capacities, and 1 Central Cloud Platform (CCP) as a national DC with big resource capacities. We consider 1008 hosting nodes distributed among these DCs offering IT resources to support the VNFs. See Fig.~\ref{fig:neo4j-psn}.

\begin{figure}[h] 
\centering
\includegraphics[width=\linewidth] {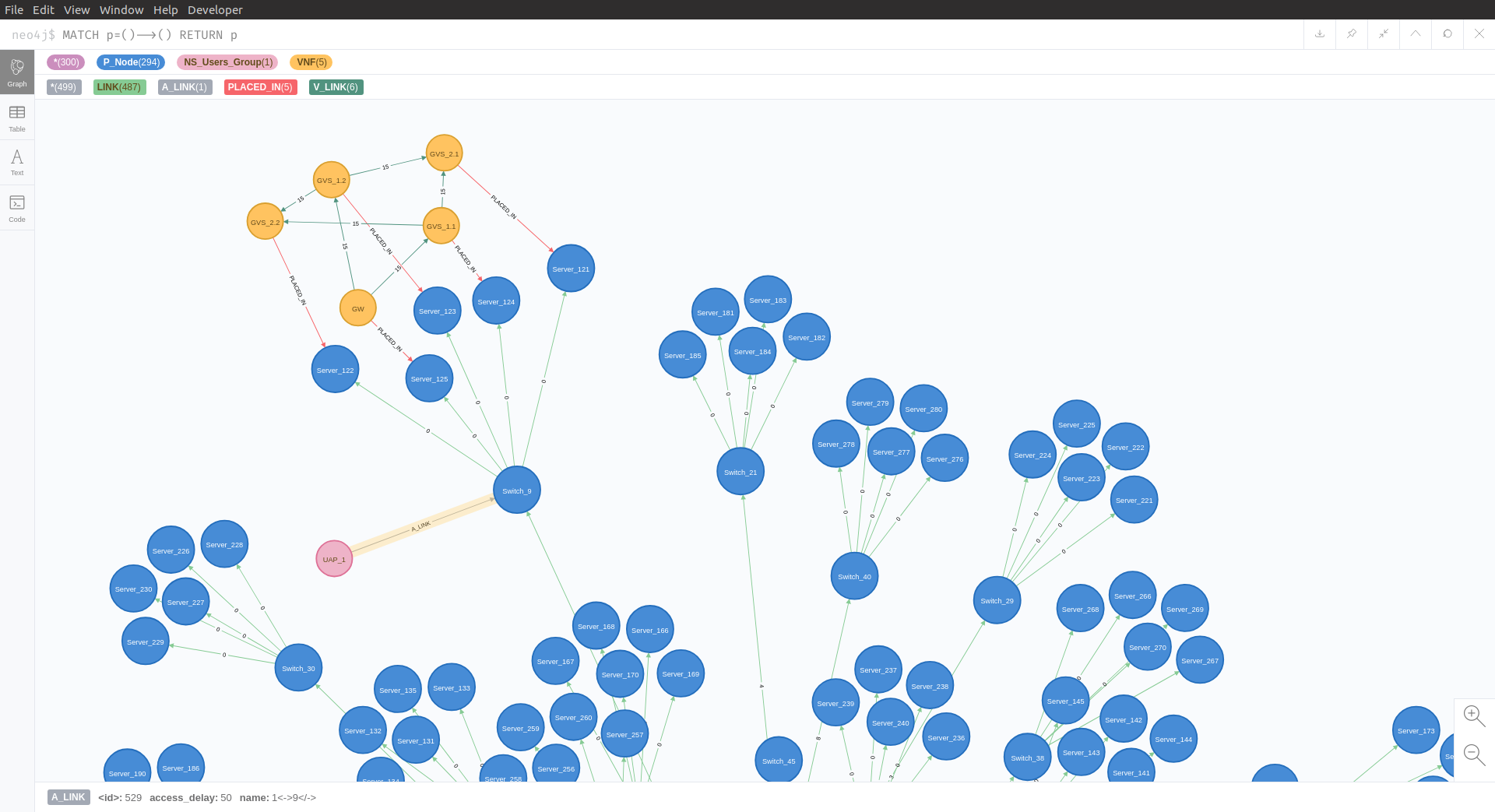}
\caption{View of a Placement Decision using the Graph visualization component}
\label{fig:neo4j-nspr}
\end{figure}

The proposed emulation is meant to demonstrate the deployment of optimized network slice placement decisions calculated with the different algorithms under a non-stationary network load scenario. PSN and NSPR are displayed by using Neo4j in Fig.~\ref{fig:neo4j-psn} and Fig.~\ref{fig:neo4j-nspr}, respectively. 

\subsection{Demonstrated Aspects}
We show 3 different aspects in our demonstration: 
\begin{enumerate}
\item the capacity of the different algorithms to learn optimal placement solutions under different network load scenarios;
\item the capability of both algorithms of sustaining a high network slice acceptance ratio when we have a critical network load,
\item how a variation of the number of training phases can impact the performance of the algorithms.
\end{enumerate}

\section{Conclusion \label{conclusion}}

This paper presents a PoC of a heuristically-controlled DRL solution for network slice placement capable of coping with  realistic and non-stationary network load conditions. We have described the implementation of the proposed solution in order to emphasize its feasibility and its accuracy when considering network slice placement in realistic non-stationary network load scenarios. All the results calculated by the placement algorithm can be easily analyzed thanks to the proposed visualization components based on Neo4j Graphs and Grafana.
As a future work, we will rely on this PoC to implement a distributed DRL algorithm for cross-domain slice placement. 

\section*{Acknowledgment}
%This paper has been conducted within 5GPPP MON-B5G project (www.monb5g.eu). 
This work has been performed in the framework of 5GPPP MON-B5G collaborative project: \textcolor{blue}{\underline{\href{https://www.monb5g.eu}{www.monb5g.eu}}}.

%José: J'ai commenté cela car EDAS m'a envoyé un erreur "URLs not allowed"

\bibliographystyle{IEEEtran}
\bibliography{my_bib}

% Generated by IEEEtran.bst, version: 1.14 (2015/08/26)
\begin{thebibliography}{1}
\providecommand{\url}[1]{#1}
\csname url@samestyle\endcsname
\providecommand{\newblock}{\relax}
\providecommand{\bibinfo}[2]{#2}
\providecommand{\BIBentrySTDinterwordspacing}{\spaceskip=0pt\relax}
\providecommand{\BIBentryALTinterwordstretchfactor}{4}
\providecommand{\BIBentryALTinterwordspacing}{\spaceskip=\fontdimen2\font plus
\BIBentryALTinterwordstretchfactor\fontdimen3\font minus
  \fontdimen4\font\relax}
\providecommand{\BIBforeignlanguage}[2]{{%
\expandafter\ifx\csname l@#1\endcsname\relax
\typeout{** WARNING: IEEEtran.bst: No hyphenation pattern has been}%
\typeout{** loaded for the language `#1'. Using the pattern for}%
\typeout{** the default language instead.}%
\else
\language=\csname l@#1\endcsname
\fi
#2}}
\providecommand{\BIBdecl}{\relax}
\BIBdecl

\bibitem{3GPP}
3GPP, ``{Management and orchestration; 5G Network Resource Model (NRM); Stage 2
  and stage 3 (Release 17)},'' {3rd Generation Partnership Project (3GPP)},
  Technical Specification (TS) 28.541, Dec. 2020, version 17.1.0.

\bibitem{etsi}
\BIBentryALTinterwordspacing
{ETSI NFV ISG}, ``{Network Functions Virtualisation (NFV); Evolution and
  Ecosystem; Report on Network Slicing Support, ETSI Standard GR NFV-EVE 012
  V3.1.1},'' ETSI, Tech. Rep., 2017. [Online]. Available:
  \url{https://www.etsi.org/technologies-clusters/technologies/nfv}
\BIBentrySTDinterwordspacing

\bibitem{vne_np_hardness}
E.~Amaldi, S.~Coniglio, A.~M. Koster, and M.~Tieves, ``On the computational
  complexity of the virtual network embedding problem,'' \emph{Electron. Notes
  Discrete Math.}, vol.~52, pp. 213--220, Jun. 2016.

\bibitem{survey_vnf_ra_2016}
J.~{Gil Herrera} and J.~F. {Botero}, ``Resource allocation in {NFV}: A
  comprehensive survey,'' \emph{IEEE Trans. Netw. Service Manag.}, vol.~13,
  no.~3, pp. 518--532, Sep. 2016.

\bibitem{p1}
Z.~{Yan}, J.~{Ge}, Y.~{Wu}, L.~{Li}, and T.~{Li}, ``Automatic virtual network
  embedding: A deep reinforcement learning approach with graph convolutional
  networks,'' \emph{"{IEEE} J. Sel. Areas Commun."}, vol.~38, no.~6, pp.
  1040--1057, Jun. 2020.

\bibitem{HA_DRL_TNSM}
J.~J. {Alves Esteves}, A.~{Boubendir}, F.~{Guillemin}, and P.~{Sens}, ``A
  heuristically assisted deep reinforcement learning approach for network slice
  placement,'' \emph{arXiv preprint arXiv:2105.06741}, 2021.

\bibitem{cnsm_2021}
J.~J.~A. Esteves, A.~Boubendir, F.~Guillemin, and P.~Sens, ``Drl-based slice
  placement under non-stationary conditions,'' \emph{Accepted to IEEE 17th Int.
  Conf. Netw. Service Manag. (CNSM)}, 2021.

\bibitem{cnsm_2020}
J.~J. {Alves Esteves}, A.~{Boubendir}, F.~{Guillemin}, and P.~{Sens},
  ``Heuristic for edge-enabled network slicing optimization using the “power
  of two choices”,'' in \emph{Proc. 2020 IEEE 16th Int. Conf. Netw. Service
  Manag. (CNSM)}, 2020, pp. 1--9.

\end{thebibliography}

\flushend

\end{document}